\documentclass[10pt]{article}

\usepackage{microtype}
\usepackage{graphicx}
\usepackage{subfigure}
\usepackage{booktabs} 
\usepackage{capt-of}

\usepackage{hyperref}

\usepackage[accepted]{mlsys2022}

\mlsystitlerunning{TESDA: Transform Enabled Statistical Detection of Attacks in Deep Neural Networks}

\usepackage{color}
\definecolor{purple}{rgb}{0.44,0.01,0.55}
\definecolor{orange}{rgb}{0.79,0.375,0.08}
\definecolor{bluey}{rgb}{0,0.48,0.65}

\usepackage{amssymb}
\usepackage{amsmath}

\begin{document}

\twocolumn[
\mlsystitle{TESDA: Transform Enabled Statistical Detection of Attacks in Deep Neural Networks}



\mlsyssetsymbol{equal}{*}

\begin{mlsysauthorlist}
\mlsysauthor{Chandramouli Amarnath}{to}
\mlsysauthor{Aishwarya H. Balwani}{to}
\mlsysauthor{Kwondo Ma}{to}
\mlsysauthor{Abhijit Chatterjee}{to}
\end{mlsysauthorlist}

\mlsysaffiliation{to}{Department of ECE, Georgia Institute of Technology}

\mlsyscorrespondingauthor{Chandramouli Amarnath}{chandamarnath@gatech.edu}

\mlsyskeywords{Deep Learning, Neural Trojans, Adversarial Attack, Attack Detection, Security}

\vskip 0.3in

\begin{abstract}
Deep neural networks (DNNs) are now the de facto choice for computer vision tasks such as image classification.
However, their complexity and ``black box" nature often renders the systems they're deployed in vulnerable to a range of security threats.
Successfully identifying such threats, especially in safety-critical real-world applications is thus of utmost importance, but still very much an open problem. 
We present TESDA, a low-overhead, flexible, and statistically grounded method for {online detection} of attacks by exploiting the discrepancies they cause in the distributions of intermediate layer features of DNNs.
Unlike most prior work, we require neither dedicated hardware to run in real-time, nor the presence of a Trojan trigger to detect discrepancies in behavior.
We empirically establish our method's usefulness and practicality across multiple architectures, datasets and diverse attacks, consistently achieving detection coverages of above 95\% with operation count overheads as low as 1-2\%.

\end{abstract}
]



\printAffiliationsAndNotice{} 

\section{Introduction}
The increasing use of intelligent systems in safety-critical applications has led to the widespread use of third-party providers to build and train Deep Neural Network (DNN) models \cite{MLaaS}.
This outsourcing, however, opens up the DNN to attack by malicious actors.
These attacks can take the form of \textbf{Neural Trojans} or \textbf{Adversarial Attacks};
In the former, backdoors are inserted into a DNN through manipulation of training, weights or data to facilitate on-line alteration of DNN behavior \cite{troj_survey}.
In the latter, inputs to a clean, non-backdoored DNN are perturbed to alter its behavior on-line \cite{advsurvey}.
We propose an approach for the detection of both, Neural Trojans and Adversarial Attacks in image classification DNNs.

\begin{figure}
    \centering
    \includegraphics[width=0.45\textwidth]{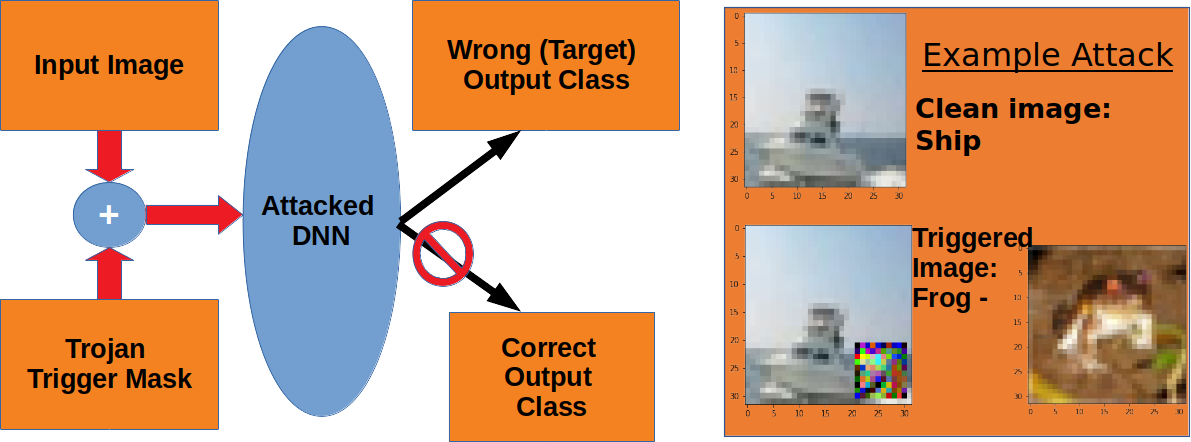}
    \caption{\textbf{\textit{Example of a Neural Trojan Attack: }}The DNN performs normally until a trigger mask is added onto the input image (boat), at which point sample is misclassified (frog).}
    \label{troj_example_cifar}
    \vspace{-0.25cm}
\end{figure}

In a \textbf{Neural Trojan}, the attacker inserts a backdoor into the DNN, called a Trojan `trigger' such that the DNN performs normally until a specific trigger mask is applied to the input image.
The trigger mask produces a specific pattern on the resultant image, causing the DNN to misclassify to a target class.
These triggers can be inserted through hardware, training data poisoning or alteration of the training process \cite{troj_survey}.
An example is shown in Figure \ref{troj_example_cifar}.

Detection of Neural Trojans is difficult due to the DNN behaving nominally \emph{until} presented with the trigger mask.
Current research either
(a) Reverse-engineers the trigger mask from a potentially infected model, typically using machine learning methods to find and patch backdoors inserted into the DNN \cite{DeepInspect, NeuralCleanse},
(b) Identifies trigger mask patterns in the input to the DNN either through input perturbation \cite{STRIP} or learning expected behavior to classify an attack \cite{Sentinet},
(c) Evaluates the training process for abnormal parameter behavior indicative of a backdoor \cite{TrojCostofClass,baracaldo2018datapoisoning},
or (d) Provides runtime detection by analyzing latent features extracted by the DNN and detecting outliers in the error of an image reconstructed from those features with respect to the input \cite{CLEANN}. 

Methods of type (a) require the DNN to be taken offline for analysis so that time-consuming machine learning methods can be applied.
Methods of type (b) incur high overhead since a compute intensive detector and DNN need to be run in sequence.
Methods of type (c) require the defender to be able to oversee the entire training process and cannot simply be used post-training.
Methods of type (d) incur high overhead due to
utilization of algorithms \cite{OMP} that cannot be parallelized, thus requiring dedicated hardware for runtime protection.
Further, none of the above methods have been tested for adversarial attack detection, and are unlikely to be able to detect adversarial attacks in the absence of trigger mask patterns.

In an \textbf{Adversarial Attack} or Adversarial Example, the attacker perturbs the inputs to the DNN to force misclassification without inserting a backdoor.
In this case, the adversarial attacks are considered to be \textit{untargeted}, aiming to force an erroneous output to any incorrect class.
Adversarial attacks typically perturb the image just enough to force a misclassification.
These attacks can be generated as an optimization problem of finding the minimally distorted image that forces misclassification \cite{advsurvey}.

Defenses against adversarial attacks on DNNs involve
(a) Adversarial training, in which attacked images are used to train the DNN \cite{PGD,YOPO},
(b) Masking the gradient used by an iterative solver to generate the minimally distorted image that forces misclassification \cite{Carlini2017b},
(c) Establishing certificates for DNN resilience to adversarial attacks through computation of upper bounds for DNN loss and modifying training to minimize them \cite{raghunathan2018},
and (d) Detect adversarial inputs prior to prediction and discard those predictions \cite{Grosse2017}.
All of (a), (b) and (c) require alteration of the training itself and thus cannot provide out-of-the-box runtime attack detection. Methods in (d) provide runtime attack detection, but have yet to be tested on Neural Trojans and can often require high-overhead machine learning methods to process the input.

We therefore present Transform Enabled Statistical Detection of Attacks (TESDA), a detection method for both Neural Trojans and Adversarial Attacks in image classification DNNs, via the extraction of pertinent features at specified intermediate layers of the DNN.
In contrast to prior art, the \textbf{key contributions} of our work are:

    \noindent(1) A \textbf{low-overhead, high-coverage real-time detection method for security attacks in DNNs}, with detection coverage of up to 99\% whilst incurring operation count overheads of only $\sim$1-2\%.\\
    \noindent(2) A flexible method for the \textbf{detection of both Neural Trojans and Adversarial Attacks} that \textit{does not rely} on static Trojan trigger mask patterns for detection.\\
    \noindent(3) \textbf{Theoretically guided hyperparameter tuning} that enables control of false positive and negative detection rates.\\
    \noindent(4) An \textbf{easy setup}, requiring only a sampling of DNN intermediate layer outputs on training data pre-deployment.


\begin{table*}[]
\centering
\caption{Qualitative Comparison of TESDA against Current State of the Art}
\vspace{0.1in}
\begin{tabular}{|l|l|l|l|l|}
\hline
\textbf{Defense Mechanism} & \textbf{On-Line Capabilities}                                              & \textbf{Overhead}                                                                    & \textbf{Tested On  }                                                                                                                           & \begin{tabular}[c]{@{}l@{}}\textbf{Trojan Trigger}\\ \textbf{Requirement}\end{tabular}           \\ \hline
CleaNN            & \begin{tabular}[c]{@{}l@{}}Detection and\\ Defense\end{tabular} & \begin{tabular}[c]{@{}l@{}}High without\\ specialized hardware\end{tabular} & \begin{tabular}[c]{@{}l@{}}(1) BadNets\\ (2) TrojanNN\end{tabular}                                                                    & Yes                                                                         \\ \hline
STRIP             & Detection                                                       & Very High                                                  & \begin{tabular}[c]{@{}l@{}}(1) BadNets\\ (2) Trojaning Attack\end{tabular}                                                            & Yes                                                                         \\ \hline
Neural Cleanse    & \begin{tabular}[c]{@{}l@{}}Detection and\\ Defense (Offline)\end{tabular} & Very High                                                                   & \begin{tabular}[c]{@{}l@{}}(1) BadNets\\ (2) Trojaning Attack\end{tabular}                                                            & \begin{tabular}[c]{@{}l@{}}No (Triggers \\ Reverse Engineered)\end{tabular} \\ \hline
TESDA & Detection                                                       & \begin{tabular}[c]{@{}l@{}}Very Low\\ (1-2\% FLOP count)\end{tabular}       & \begin{tabular}[c]{@{}l@{}}(1) Targeted Bit\\ Trojan\\ (2) Input-Aware \\ Backdoor Attack\\  (3) PGD Adversarial\\ Attack\end{tabular} & Not Necessary                                                               \\ \hline
\end{tabular}
\label{comparisontable_intro}
\end{table*}

The paper is organized as follows: Section \ref{Background} discusses prior art, contrasting our choice of test case attacks and assumptions versus those of previous methods.
Sections \ref{Overview}, \ref{Details}, \ref{Experimental} present an overview of our proposed approach, its detailed description, and our experimental results on commonly used DNNs and datasets respectively.
Sections \ref{Discussion}, \ref{Conclusion} end with a discussion of some ablation studies and conclusions.


\section{Background}\label{Background}
\subsection{Threat Models and Approach Assumptions}

The three attacks tested here are discussed below. An overview of the assumptions made by the proposed detector is then compared qualitatively against prior art.

    \noindent\textbf{(1) Targeted Bit Trojan: }The Targeted Bit Trojan (TBT) attack \cite{TBT} creates DNN backdoors via bitflips made in layer weights stored in memory. The trigger mask is a static pattern. The attack is inserted \textit{after} DNN deployment. The attacker is assumed to have access to a subset of training data and a copy of the trained model to build the Trojan and to insert it via bitflips in weights.\\
    \noindent\textbf{(2) Input-Aware Backdoor Attack: }In this attack \cite{InputAwareBackdoor} the DNN a backdoor is inserted during training such that the DNN misclassifies to a target class when presented with the trigger mask. This attack is different from prior art due to its use of a \textit{dynamic} trigger mask as a function of the input. The attacker is assumed to have control of the training process, shipping a contaminated DNN around which our proposed detector will be built.\\
    \noindent\textbf{(3) Projected Gradient Descent (PGD) Attack: }The attacker is assumed to have information about the trained model and its gradients to compute the input perturbations to force DNN misclassification to any class \cite{PGD}. The PGD attack accomplishes this by attempting to find the input perturbations that maximize DNN loss while keeping perturbation size smaller than a predefined bound.

In contrast to prior art, the proposed detector is built around a trained DNN model and requires access to the trained DNN's intermediate layer outputs on clean training data to be able to profile the DNN. The detector can be built on a clean or contaminated model, does not require DNN re-training and is not tailored to specific attacks.

\subsection{Comparison to Prior Art}

A qualitative comparison of the proposed approach against prior art is provided in Table \ref{comparisontable_intro}. We compare against CleaNN \cite{CLEANN}, STRIP \cite{STRIP} and Neural Cleanse \cite{NeuralCleanse}. STRIP provides on-line detection of Neural Trojans but incurs very high overhead due to the need to repeatedly perturb inputs to detect a Trojan trigger. CleaNN provides real-time detection and defense against Trojan triggers via dedicated FPGA hardware acceleration but has high overhead without that hardware. Neural Cleanse uses machine learning methods to find and patch backdoors in a DNN and does not provide real time attack detection. In contrast to this, TESDA provides low-overhead real-time attack detection with operation count overheads of 1-2\% before acceleration or optimization and does not require dedicated hardware. TESDA does not use iterative methods leveraged by prior art such as OMP \cite{OMP}, thereby enabling future parallelization.

All other approaches in Table \ref{comparisontable_intro} have been tested by their authors against BadNets \cite{BadNets}, an attack which inserts Trojan backdoors during training. Neural Cleanse and STRIP are also tested by their authors against a backdoor Trojaning Attack \cite{LiuBackdoor}. CleaNN is also tested by its authors against the backdoor attack of TrojanNN \cite{TrojanNN}. By contrast, our approach is verified on the input-aware dynamic backdoor attack \cite{InputAwareBackdoor}, which has been shown to break STRIP and Neural Cleanse using input-specific trigger masks. TESDA is also verified on the Targeted Bit Trojan \cite{TBT}, which is inserted after training, bypassing offline defenses. The greater flexibility of TESDA is verified by tests on the Projected Gradient Descent adversarial attack \cite{PGD}. Our approach does not require static trigger mask patterns to detect an attack.

\section{Approach Overview}\label{Overview}

\subsection{Network-Level Detector Deployment}\label{network}

\begin{figure}
    \centering
    \includegraphics[width=0.48\textwidth]{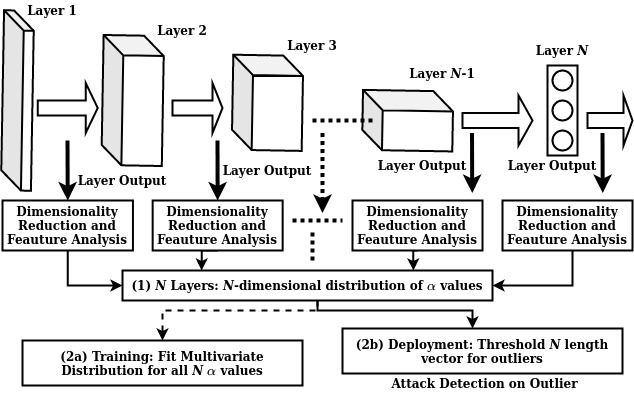}
    \caption{\textbf{\textit{Attack Detection Process for a DNN: }}Layer outputs are reduced to features ($\alpha$s). Statistical tests on the feature vector detect outliers indicating an attack. Further details in Section \ref{network}.}
    \label{fig:DNN_flow}
    \vspace{-0.55cm}
\end{figure}

Figure \ref{fig:DNN_flow} shows the deployment of TESDA on a DNN. Each layer's output is reduced to a feature coefficient $\alpha$ in the dimensionality reduction and feature extraction blocks. These blocks employ identical methods for each layer and are discussed further in the following subsection. For a network of $N$ layers this yields an $N$ length vector of $\alpha$s, $\theta$ in Block 1. A multidimensional distribution is fitted to the set of $\theta$s across the DNN training data to enable attack detection (Block 2a). If the $\theta$ produced during operation is an outlier with respect to that distribution (Block 2b), it is considered indicative of an attack. 

\subsection{Detector Components: Single Layer}\label{singlelayer}

\begin{figure}
    \centering
    \includegraphics[width=0.48\textwidth]{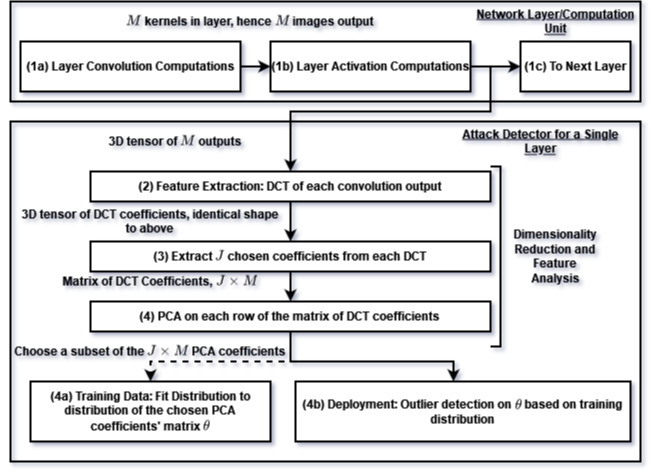}
    \vspace{-0.2cm}
    \caption{\textbf{\textit{Attack Detection Process for a single layer: }}The feature extraction (DCT, PCA) are used by every layer of Figure \ref{fig:DNN_flow} before statistical tests for attack detection.}
    \label{fig:singlelayer_flow}
    \vspace{-0.25cm}
\end{figure}

Figure \ref{fig:singlelayer_flow} shows the process of attack detection applied to a single convolutional layer of $M$ kernels.
After convolution, bias addition and activation, each kernel produces a 2D image, giving a layer output of an $M$ such images, a 3D tensor. The size of this tensor can make direct analysis expensive. The first step is to reduce this to low dimensional features. This is done in blocks 2-4 of Figure \ref{fig:singlelayer_flow}, which takes the layer output from Block 1b.


The first step in feature extraction and dimensionality reduction is reduction of each kernel output to its frequency coefficients using the Discrete Cosine Transform (DCT) \cite{DCT} (Block 2). This gives an $M$-length transformed tensor where each element is a 2D array of DCT frequency coefficients. One or more of the DCT coefficients from each transformed kernel output may be chosen for further analysis. In Block 3 of Figure \ref{fig:singlelayer_flow}, $J$ DCT coefficients were chosen from each element of the tensor, giving an $J\times M$ array of DCT coefficients.

In Block 4, a Principal Components Analysis (PCA) \cite{PCA} transformation is applied to each of the $J$ frequency coefficients in the $M$-length array obtained from Block-3. The PCA is fit to each row of the set of chosen DCT coefficients from the layer outputs across the training data set. This yields an $J\times M$ array of projections, with coefficients $T=[\alpha_1,\alpha_1,...,\alpha_{M}]$ ordered according to significance. Each $\alpha_i \in \mathbb{R}^J$ is a vector of the $i$th PCA coefficients of all $J$ DCT coefficients chosen. One or more $\alpha$s can be chosen from $T$ for attack detection in Blocks 4a and 4b. For a dense layer, a PCA is taken along the layer output vector directly and a subset of $T$ can be used for attack detection.

Attack detection is done based on the distribution of the chosen subset of $T$, $\theta$. If the $\theta$ from Block-4 is an outlier with respect to its distribution over training data, an attack is indicated. A multivariate distribution is fitted to the $\theta$ values on the training data in Block 4a. This distribution is used to set thresholds for outlier detection post-deployment of the DNN and detector. Statistical tests leveraging the training distribution for outlier detection are used here, allowing threshold adjustment to ensure a desired false positive rate.

\section{Approach Details}\label{Details}
\subsection{Feature Extraction and Dimensionality Reduction}

In Blocks 2-4 of Figure \ref{fig:singlelayer_flow} the output tensor of each convolutional layer under examination is reduced to a set of PCA coefficients $\theta$ that represents layer behavior. This begins with the Discrete Cosine Transform of Block-2 to extract the frequency components of each convolution kernel output.

\subsubsection{DCT Analysis}

Each convolution kernel in each layer under analysis outputs an image after activation. The DCT \cite{DCT} is taken on the output of each kernel to reduce these images to frequency components in the form of a sequence of sinusoids (cosines) in two dimensions oscillating. A set of $J$ frequency components is chosen for further analysis. The same components are chosen across all transformed kernel outputs. This yields a matrix of the chosen DCT coefficients $D_i$ for the $i$th layer in the network. While Trojan trigger presence has been seen to typically affect higher frequency coefficients \cite{CLEANN}, the flow in Figure \ref{fig:singlelayer_flow} also targets adversarial attacks. A range of DCT coefficients is thus tested in this work. In our experiments we predominantly use $J=1$. Further experiments using $J>1$ are detailed in Appendix \ref{appendix_dct_coeffs}.

\subsubsection{Principal Component Analysis}\label{PCA_subsec}

For a network of $N$ layers, the process above yields a set $D = \{D_i\}_{i=1}^N$ where each $D_i \in \mathbb{R}^{J \times M_i}$ is a matrix of DCT coefficients.
We further reduce the dimensionality of this set using Principal Component Analysis (PCA) \cite{PCA}, by applying it independently to each $D_i$. 

The PCA transforms each row of $D_i$ to a vector space whose basis vectors are orthogonal to one another. The elements of each row of the transformed matrix $T_i$ are ordered according to the energy in each principal component (or basis vector of the transform space), providing an optimal decomposition of each row of $D_i$. As a linear transformation, the PCA has low overhead. The transformation matrices for the $i$th layer are obtained from the mean and covariance of the rows of the set of $D_i$ produced across the training data set.

\begin{figure}
    \centering
    \includegraphics[width=0.45\textwidth]{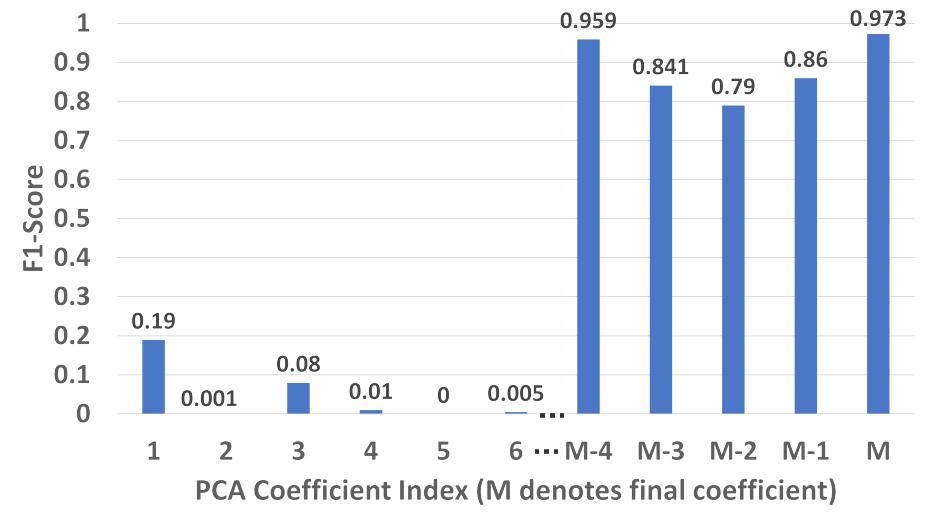}
    \caption{\textbf{\textit{Outlier Detection Performance for Singular PCA Coefficients: }}
    The detection f1 scores when using one of the first 6 or last 5 ($M-4,...M$)
    PCA coefficients of a simgle DNN layer are shown above.
    The trends clearly reflect that higher energy coefficients have low detection performance while the high energy ones have are better suited for the task.}
    \label{fig:alphas_choice}
    \vspace{-0.25cm}
\end{figure}

The vector of DCT coefficients is thus transformed into a matrix $T_i \in \mathbb{R}^{J\times M_i}$ for each layer $i$, $1\leq i\leq N$ such that each column of $T_i$, $\alpha_k \in \mathbb{R}^J, 1\leq k\leq M_i$ is a vector of the $k$th PCA coefficients of all $J$ DCT coefficients chosen. For a dense layer the DCT decomposition is not required and the PCA transform is taken from the output vector of the dense layer for each input to the DNN. The PCA for a dense layer is fitted to the set of outputs produced for each input in the training data set.

One or more columns can then be taken from each $T_i$ for the purposes of outlier detection, reducing the high-dimensional tensor output from a convolutional layer to just a vector containing few coefficients.
We note however that the higher energy coefficients (e.g., $\alpha_1, \alpha_2$) are seen to contain lesser information about the outliers, i.e., perform worse at outlier detection (Figure \ref{fig:alphas_choice}) than their lower energy counterparts (e.g., $\alpha_{M_i}$).
We therefore restrict ourselves to using just the lowest energy coefficient corresponding to each layer, giving us a vector $\theta \in \mathbb{R}^{N}$ for every sample, where $\forall i \in \{1,2,...N\},\; \theta_i = \alpha_{M_i}.$

\subsection{Attack Detection with Elliptic Envelopes}
\label{attack_detection}

\begin{figure}
    \centering
    \includegraphics[width=0.45\textwidth]{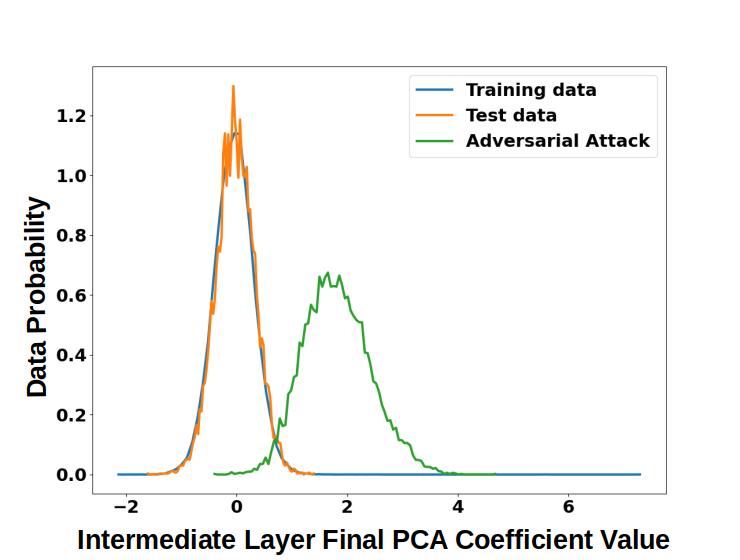}
    \caption{\emph{\textbf{Distributions of $\alpha$ for clean and corrupted samples.}} Probability density functions of the last $\alpha$ for features at the second residual block of a ResNet-18 on the CIFAR-10 dataset. The adversarial attack dramatically shifts both the mean and variance of the clean distribution.
    The curves have been re-normalized such that the area under each is unity.} 
    \label{fig:example_distribs}
    \vspace{-0.25cm} 
\end{figure}

To detect an attack we check if the $\theta$ obtained for a particular sample is an \emph{outlier} with respect to the true (i.e., clean) distribution of $\theta$ as observed in the training set.
This of course would only work if an attack appreciably alters the distribution of the $\alpha$ used in construction of the vector $\theta$, which we do confirm empirically to hold true (Figure \ref{fig:example_distribs}).

\begin{figure}
    \centering
    \includegraphics[width=0.4\textwidth]{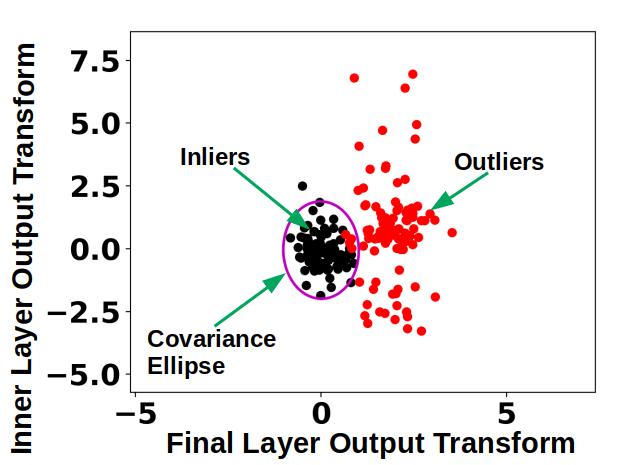}
    \caption{\emph{\textbf{Example Outlier Detection for ResNet-18.}} A Multidimensional Gaussian is fit to the $\theta$ produced by features from the second residual block a ResNet-18 trained on the CIFAR-10 dataset.
    Clean samples are shown in black and corrupted samples are shown in red.
    The presence of any outliers is assumed to be indicative of abnormal behavior, i.e., an attack.
    }
    \label{fig:example_ellipse}
\end{figure}

We then approximate the distribution of the clean $\theta$ across the training set as a \textbf{multidimensional Gaussian} $\mathcal{N}(\mu,\Sigma)$, allowing us to leverage the statistical tools available for robust Gaussian parameter estimation and model fitting.
Specifically, we use the minimum covariance determinant (MCD) method \cite{rousseeuw1984least, rousseeuw1985multivariate, covarianceellipse} which given a training set with $n$ $k$-dimensional samples of which at most $m$ are outliers, finds estimates $\hat{\mu} \textrm{ and } \hat{\Sigma}$ for the true mean and covariance.
To do so MCD finds $h = [\frac{n+k+1}{2}]$ samples from the training set whose covariance matrix has the least determinant and fits a Gaussian to them, implicitly making the assumption that $m \lessapprox n - h = \frac{n-k-1}{2}.$
Subsequently, any sample that is \emph{outside} the \textbf{Elliptic Envelope} described by $\hat{\mu}, \hat{\Sigma}$, i.e., has a {Mahalanobis distance} $> \Delta$, a
pre-specified threshold, is classified as an outlier.
It should be noted that for Gaussian models the MCD method has been shown to be asymptotically consistent \cite{butler1993asymptotics}, i.e., as $n \rightarrow \infty, \hat{\mu} \rightarrow \mu \textrm{ and } \hat{\Sigma} \rightarrow \Sigma$.
An example of the MCD based elliptic envelope fit to noisy data processed through a DNN is shown in Figure \ref{fig:example_ellipse}.

\subsubsection{Hyperparameter Tuning to Match Target False Positive and False Negative Rates}
\label{hyperparam_tuning_for_fnr_fpr}

Due to its simplicity and interpretability, the MCD method lends itself to principled hyperparameter tuning which allows us to choose the offset $\Delta$ such that given an approximate upper bound on the degree of contamination of the training data, $\varepsilon \in (0,0.5)$, we can stay under either a target false negative (FNR) or false positive rate (FPR).

Let $\hat{\mu} \in \mathbb{R}^k \textrm{ and positive definite } \hat{\Sigma} \in \mathbb{R}^{k \times k}$ be the robust sample mean and covariance estimates
provided by the MCD method.
Further assume $n > h \gg k$ and $\hat{\mu} \approx \mu, \hat{\Sigma} \approx \Sigma.$
For a sample $\theta$ to be classified as an outlier (i.e., an attack) we require that $d \geq \Delta$, or equivalently $d^2 > \Delta^2$, where $d^2 = {(\theta-\hat{\mu})^T\hat{\Sigma}^{-1}(\theta-\hat{\mu})}$ is the squared Mahalanobis distance of $\theta$ and $\Delta$ is a constant computed over and fit to the entire training data.
Specifically, $\Delta$ depends on the contamination parameter $\varepsilon$ and is fit on the training data such that the number of outliers $m \approx \varepsilon n$.

Given that $\varepsilon$ acts as an upper bound on the contamination of the training set, we want to pick $\Delta^2$ such that $\mathbb{P}[d^2 \geq \Delta^2] \leq \varepsilon$.
At the same time, by the multivariate Chebyshev inequality \cite{stellato2017multivariate} we have $\mathbb{P}[d^2 \geq t^2] \leq \frac{k(n^2 - 4 + 2nt^2)}{n^2t^2}$.
Equating the two, we get $\Delta = \sqrt{\frac{k(n^2-4)}{\varepsilon n^2 - 2nk}}$.

Provided an expression for $\Delta$ we have a principled way of choosing the contamination parameter $\varepsilon$ so as to stay under a specified false negative ($\tau_N$) or false positive ($\tau_P$) rate.

\textbf{False negative rate}: The FNR is by definition the complement of the true positive rate, which on the training set must be $\varepsilon$.
Therefore to minimize FNR, we set $\varepsilon = 1 - \tau_N$ and then fix $\Delta \leq{\sqrt{\frac{k(n^2-4)}{\varepsilon n^2 - 2nk}}} = {\sqrt{\frac{k(n^2-4)}{(1-\tau_N)n^2 - 2nk}}}$.

\textbf{False positive rate}: To minimize false positives we simply require that for clean samples $\mathbb{P}[d^2 \geq \Delta^2] \leq \tau_P$.
We therefore set $\varepsilon = \tau_P$ and fix $\Delta \geq {\sqrt{\frac{k(n^2-4)}{\varepsilon n^2 - 2nk}}} = {\sqrt{\frac{k(n^2-4)}{(\tau_P)n^2 - 2nk}}}.$

Additional details regarding the derivation of the expression for $\Delta$ are provided in Appendix \ref{tuning_delta_details}. 

\subsubsection{Deriving Tighter Bounds for Outlier Detection}

Leveraging the fact that the distribution of $d^2$ for clean $\theta$ is given by the chi-squared distribution with $k$ degrees of freedom, i.e., $\chi_k^2$ (Appendix \ref{mahalanobis_chi_squared}), it is possible to have bounds tighter than those using Chebyshev's inequality, with the caveats that
i) the expression for $\Delta$ would have to be split into multiple expressions depending on its range,
or
ii) the dependence of $\Delta$ on $\varepsilon$ might no longer be expressible as closed form expressions of elementary functions.

For example, noting that the distribution $\chi^2_k$ is sub-exponential with parameters $(2k,4)$ \cite{ghosh2021exponential}, one can write the corresponding sub-exponential tail bound \cite{wainwright2019high}, resulting in the following expressions
\begin{equation*}
\Delta = \begin{cases}
\sqrt{4k\sqrt{\ln(\frac{1}{\varepsilon^2})}+k}\; &\sqrt{k} \leq \Delta \leq \sqrt{k^2+k}\\
\sqrt{8\ln(\frac{1}{\varepsilon})+k}\; &\Delta > \sqrt{k^2+k}
\end{cases}
\end{equation*}
Alternatively, one could also write a Chernoff bound \cite{wainwright2019high} that yields $\Delta = \sqrt{-kW(\frac{-\varepsilon^{2/k}}{e})}$, $W$ being the Lambert \emph{W} function \cite{bronstein2008algebraic} whose values may be calculated numerically. 

Given an expression for $\Delta$, setting $\varepsilon = 1-\tau_N$ (or $\varepsilon = \tau_P$) results in the value of $\Delta$ that matches a target FNR (or FPR) as described in Section \ref{hyperparam_tuning_for_fnr_fpr}.
Details of the derivations of the expressions for $\Delta$ corresponding to the tail bounds mentioned above are provided in Appendices \ref{subexponential_delta_bounds} and \ref{chernoff_delta_bounds}.

\subsection{The TESDA Algorithm}

We consolidate and present the complete detection process in Algorithm \ref{alg:detector}.
TESDA begins by taking as input the $N$ layer wise outputs from the network $[L_1,L_2,..,L_N]$ (Line 1) of a DNN that produces a label $Y$ for an input $X$.
It then initializes the array for the input $\theta$ to the outlier detector (Line 2) and calls the processing loop (Line 3) that runs for each of the $N$ layers.
In the loop, first, DCT is applied to the outputs of each kernel of the $i$th convolutional layer and DCT-Analyzer() returns the matrix of $J\times M_i$ selected DCT coefficients $D_i$ (Line 4). In case of a dense layer, the DCT step is skipped and the function simply returns the vector of layer outputs.
Next, the rows of $D_i$ over the entire training data are transformed using PCA (Line 5) to produce a matrix $T_i$ of PCA coefficients, where every row of $T_i$ corresponds to a layer of the DNN and every column corresponds to a PC.
Starting with the last one, columns of $T_i$ are assigned as features of the vector $\theta$ (Line 6) and the complete $\theta \in \mathbb{R}^{JN}$ is fed to Elliptic-Envelope, i.e., the outlier detector (Line 8).
If $\theta$ is classified as an outlier, we flag it as an attack on the network.
We empirically validate TESDA on a variety of test cases, the results of which are presented in Section \ref{Experimental}.



\begin{algorithm}[h]
   \caption{TESDA Algorithm}
   \label{alg:detector}
\begin{algorithmic}[1]
   \STATE $Y$, $[L_1,L_2,...,L_N]\leftarrow$ DNN($X$)
   \STATE Initialize Empty $\theta$
   \FOR{i=1 {\bfseries to} $N$}
    \STATE $D_i\leftarrow$ DCT-Analyzer($L_i$)
    \STATE $T_i\leftarrow$ PCA($D_i$)
    \STATE $\theta [i]\leftarrow T_i[-1]$
   \ENDFOR
   \STATE Detection $\leftarrow$ Elliptic-Envelope($\theta$)
\end{algorithmic}
\end{algorithm}

\section{Experimental Results}\label{Experimental}
\subsection{Experimental Metrics}\label{metrics}

We evaluate TESDA using the following metrics:\\
\textbf{Detection Coverage:} Detection coverage is the number of detected attacks divided by the total number of attacks. This is the \textit{true positive rate} of the attack detection system.\\
\textbf{False Positive Rate (FPR): }This is the percentage of time that the detector flags clean operation, obtained by running the network and detector across clean test data. \\
\textbf{F1-Score:} The F1-score is the harmonic mean of precision and recall. Precision is the true positive rate divided by the sum of true and false positive rates. Recall is the true positive rate divided by the sum of true positive and negative rate. This is represented as a percentage value from 0-100\%.

\textit{Algorithm overhead} for the attack detector is calculated using CPU performance counter data and measured relative to the resources consumed by the DNN run on the CPU. Further details are provided in Section \ref{HardwareOverhead}.

\subsection{Network, Dataset and Ablation Details}\label{networks}

The datasets used here are CIFAR-10 \cite{cifar10} and GTSRB \cite{GTSRB}. CIFAR-10 consists of 32x32 color images, with 50K training and 10K test images across 10 classes. GTSRB consists of 43 traffic sign classes across a training set of 39209 labeled images and 12630 test images. Tests were conducted on these for two DNNs.

The \textit{first} network tested is \textit{ResNet-18} \cite{Resnet} using the CIFAR-10 dataset. It has one input convolutional layer, four \textit{residual blocks}, one average pooling and one output linear layer. In each \textit{residual block} an identity mapping combines its input with the result of a convolution `branch' for ReLu activation. The \textit{second} network examined is \textit{PreAct ResNet-18} \cite{PreactResNet} using the GTSRB dataset. For PreAct ResNet, the `branch' is identical to ResNet-18, with one ReLu moved from the end of the unit to the start of the branch. PreAct ResNet-18 has one convolutional input layer, four \textit{residual blocks}, one average pooling layer and one output linear layer. The \textit{detector} for both networks conducts feature extraction with $J=1$ and selects the final $\alpha_{M_i}$ from the PCA matrix $T_i$ at each residual block for detection.

\noindent\textbf{\textit{Ablation Test Cases: }}The performance of the system is tested for (1) Different outlier thresholds $\varepsilon$ with the detector concurrently connected across all residual blocks and the final layer for ResNet-18 and PreAct ResNet-18 and (2) For each residual block and the final layer with the detector connected to them \textit{individually}. Metrics are recorded as per Section \ref{metrics}. For relevant test cases we also record (3) The effects of using different single DCT coefficients.

\subsection{Targeted Bit Trojan}

The Targeted Bit Trojan (TBT) attack \cite{TBT} creates backdoors in a DNN, forcing misclassification when presented with the trigger mask via bitflips made in layer weights stored in memory. The attack was generated by modifying code provided by the authors.

The attack works as follows: (a) In the \textit{Insertion} step, alterations to specific weight bits are made to map the input to a specific target class in the presence of the trigger. The trigger is designed to force vulnerable neurons in the target layer linked to the target class to fire at a large value. The TBT flips vulnerable layer weight bits in memory such that the DNN performs nominally without the trigger. It is shown in \cite{TBT} that only 84 bit flips out of 88 million bits are needed to classify 92\% of images to a target class on Resnet-18 for CIFAR10. Since these weights are stored in memory, row-hammer attacks \cite{rowhammer_kim} can be used for insertion. (b) In the \textit{triggering} step of the attack, the image trigger (mask) is inserted into the DNN input image. The mask modifies pixel values in a small area of the image and causes mis-classification by the DNN to an incorrect target class, as in Figure \ref{troj_example_cifar}. The TBT is thus inserted \textit{after} deployment of the DNN. This bypasses pre-deployment scans. The TBT here attacks the \textit{final} layer of ResNet-18, forcing misclassification to target class 2 (Bird).

TESDA is tested for the TBT using a pretrained ResNet-18 trained on CIFAR-10 provided by the authors for cases (1) and (2) from Section \ref{networks}. Since the TBT only attacks the final linear layer, there are no material changes to convolutional layers and case (3) is not evaluated (the DCT is taken only for convolutional layers).

\begin{figure}
    \centering
    \includegraphics[width=0.5\textwidth]{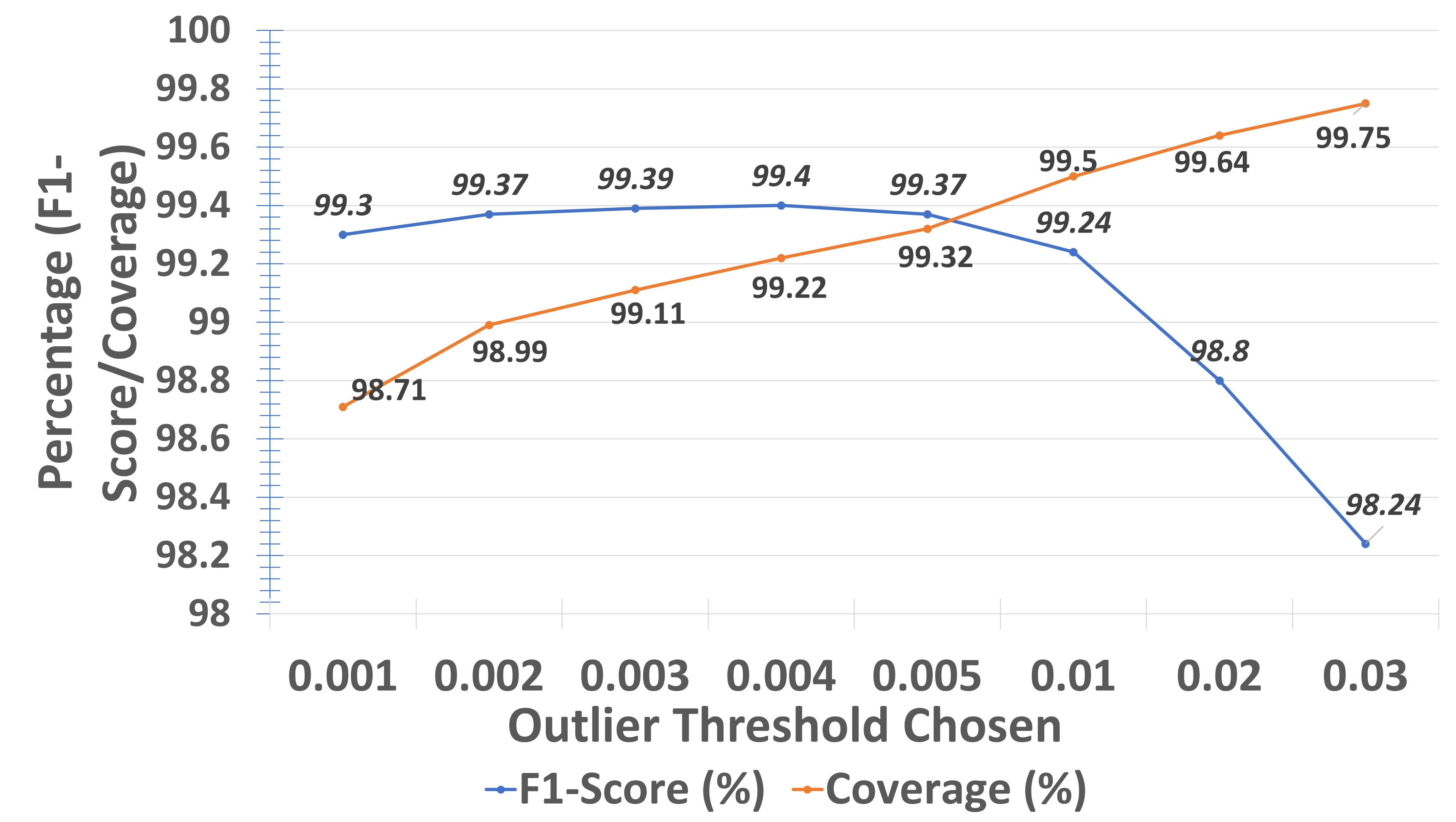}
    \caption{\textbf{\textit{Detection Performance Against TBT on ResNet-18, Varying Thresholds: }}The detector is connected concurrently on all four residual blocks and the final linear layer.}
    \label{fig:TBT_thresh}
    \vspace{-0.25cm}
\end{figure}

(1) Detector performance for different levels of outlier threshold $\varepsilon$ is shown in Figure \ref{fig:TBT_thresh}. By varying $\varepsilon$, the level of false positives seen in the detector can be adjusted.  For all outlier thresholds, detection coverage remains above 99\% and varies only marginally, while F1-score falls as the threshold rises. Italicized figures are data point values for F1-score and non-italicized ones are for coverage.

\begin{figure}
    \centering
    \includegraphics[width=0.5\textwidth]{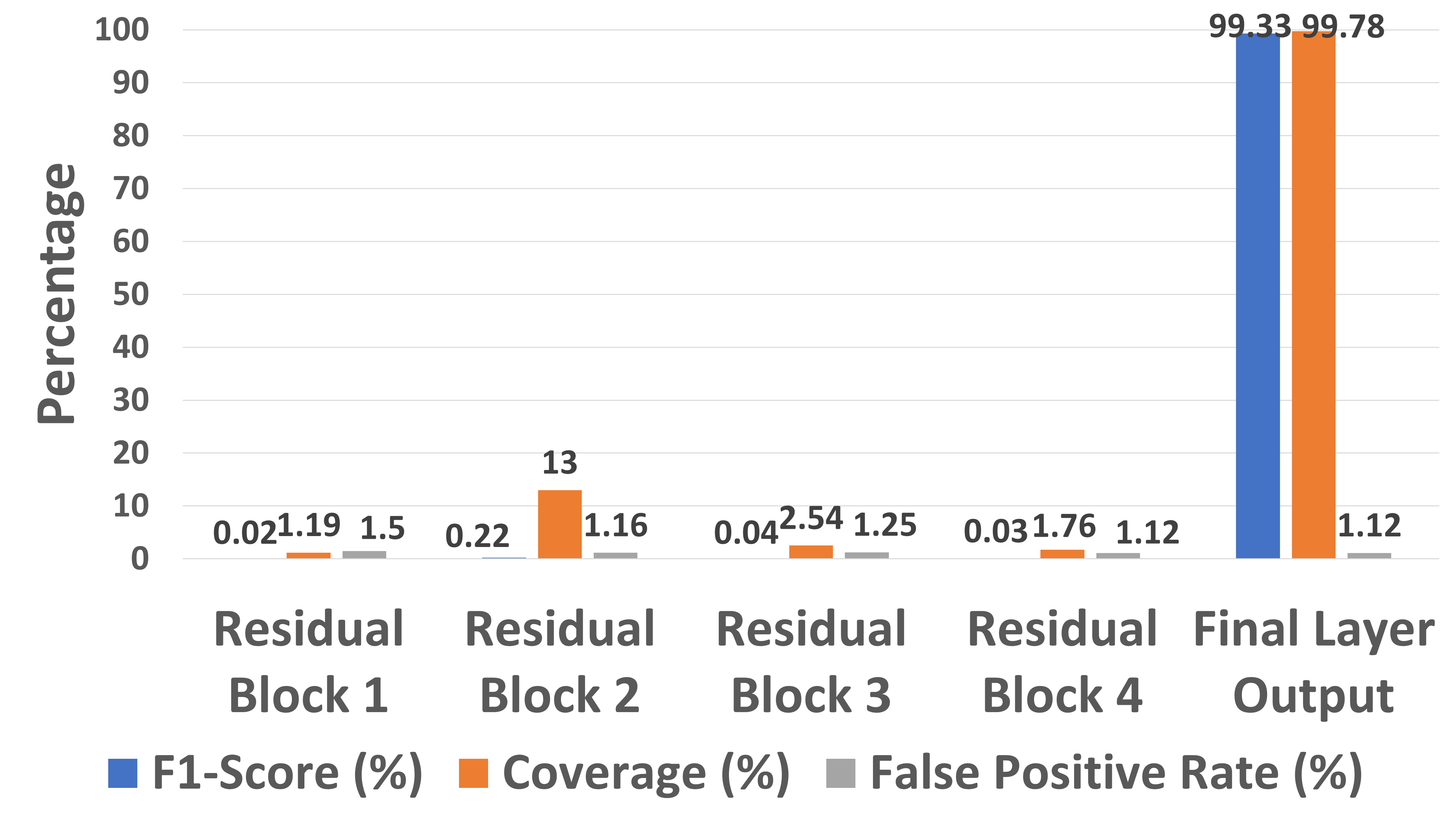}
    \caption{\textbf{\textit{Detection Performance By Layer Against TBT: }} The detector is connected across each residual block and the final layer of ResNet-18 \textit{individually}}
    \label{fig:TBT_Layer_Ablation}
    \vspace{-0.25cm}
\end{figure}

(2) Detector performance for each residual block and the final layer is shown in Figure \ref{fig:TBT_Layer_Ablation}. The detector is connected across each unit and the final layer alone. Clean and Trojaned data are run across the detector. The F1-score is negligible until the final layer, which is the one targeted by the TBT. Here the F1-score and the coverage remain above 99\%. $\varepsilon$ used here was 0.01.
\subsection{Input-Aware Dynamic Backdoor Attack}

The second attack tested is the Input-Aware Dynamic Backdoor Attack applied to PreAct ResNet-18 on GTSRB using code provided by the authors \cite{InputAwareBackdoor}. In this attack, a backdoor is inserted during DNN training to force misclassification to a target class when presented with the trigger pattern. This attack differs from prior benchmarks in its use of an \textit{input-aware} dynamic trigger mask. The trigger pattern generated in training is a function of the input image, dictated by an autoencoder. Training is modified to ensure the triggers are usable only for the image they are intended for. These \textit{dynamic triggers} have been shown to bypass defenses such as STRIP. The attack has more than 99\% effectiveness on GTSRB. The trained DNN is already contaminated when run on clean training data to set up the detector. The detector set up based on the $\theta$ training set generated by this contaminated network is used to detect the attack when a trigger mask is passed. The defender is assumed to be able to run the contaminated DNN on clean training data before deployment. The attack here forces misclassification to target class 0 (20kph speed limit sign).

The detector performance for this attack is tested on PreAct ResNet-18 for cases 1-3 of Section \ref{networks}. 

\begin{figure}
    \centering
    \includegraphics[width=0.5\textwidth]{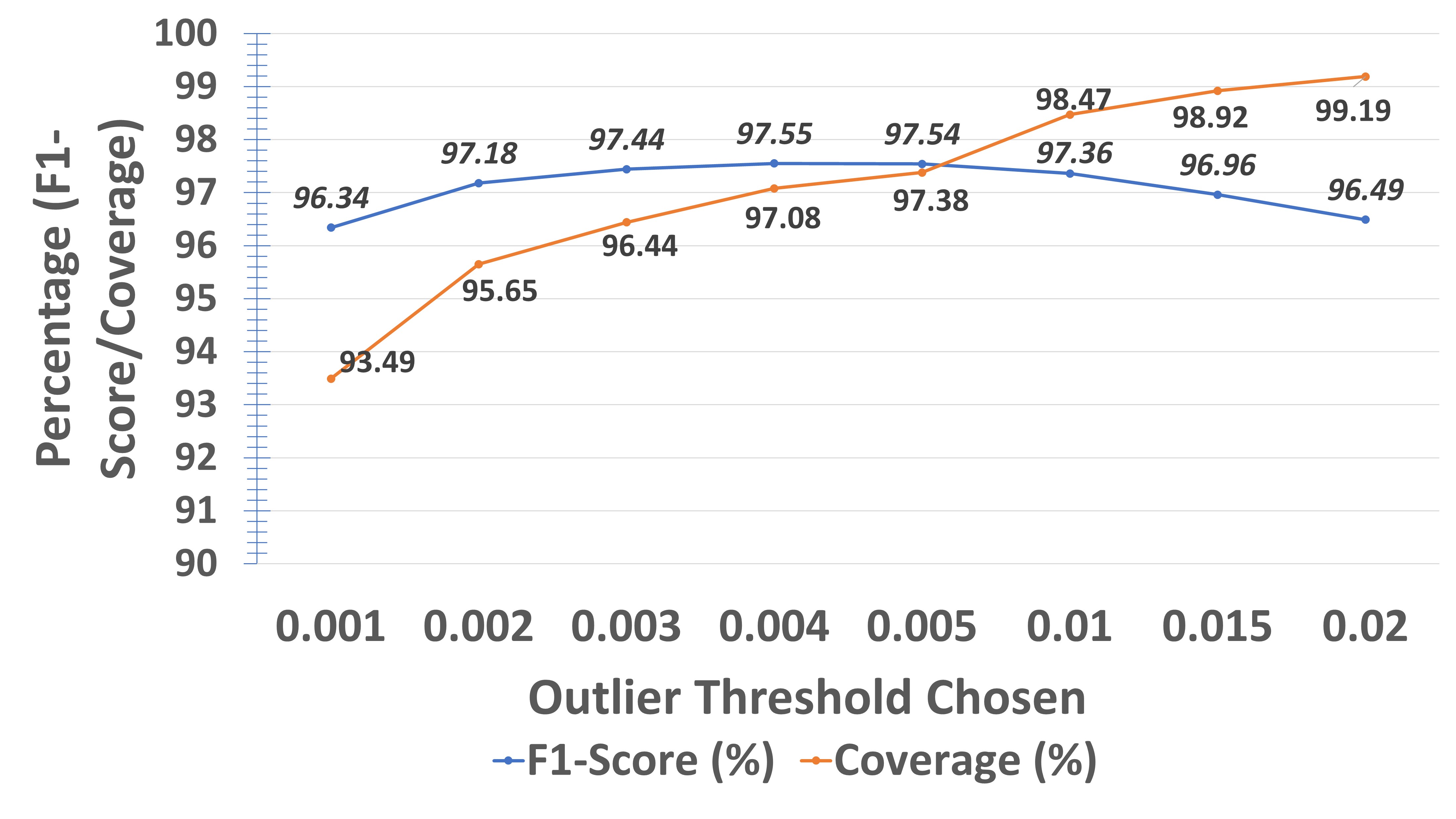}
    \caption{\textbf{\textit{Detection Performance Against Input-Aware Backdoor on PreAct Resnet-18, Varying Thresholds: }}The detector is connected concurrently on all four residual blocks and the final linear layer.}
    \label{fig:Thresh_Backdoor}
    \vspace{-0.25cm}
\end{figure}

\begin{figure}
    \centering
    \includegraphics[width=0.5\textwidth]{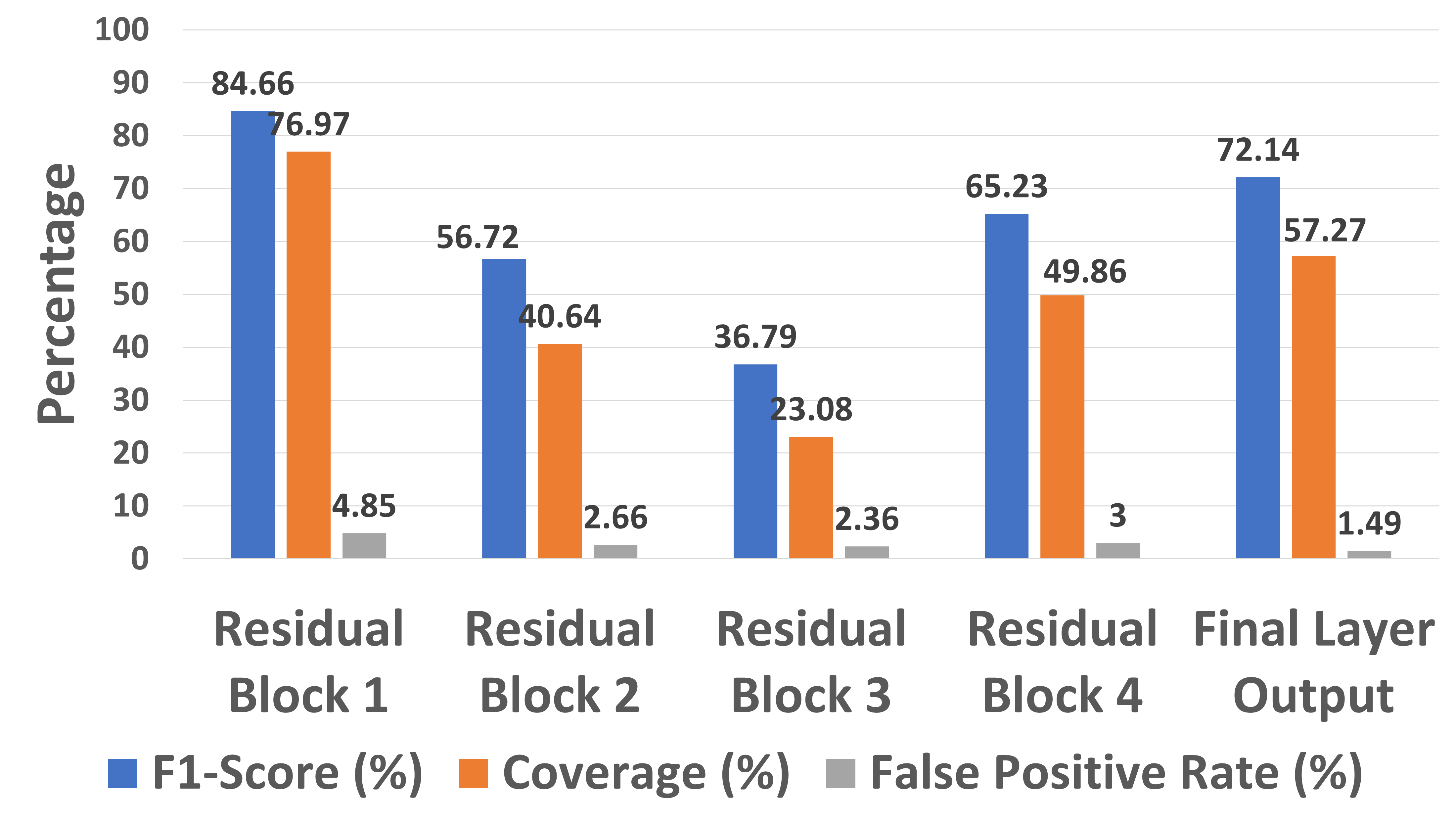}
    \caption{\textbf{\textit{Detection Performance By Layer Against Input-Aware Backdoor: }}The detector is connected across each residual block and final layer of PreAct ResNet-18 \textit{individually}}
    \label{fig:Layer_Ablation_Backdoor}
\end{figure}

(1) Detector performance for different levels of outlier detection threshold $\varepsilon$ is shown in Figure \ref{fig:Thresh_Backdoor}. The italicized figures are the data point values for F1-score and the non-italicized ones are for the coverage. The F1-score peaks at $\varepsilon=0.004$ or a 0.4\% expected false positive rate. This gives a coverage of 97.08\% and an FPR of 1.95\%. A threshold of $\varepsilon=0.01$ yields a coverage of 98.4\% and a false positive rate of 3.8\%. 

(2) Detector performance for each residual block is shown in Figure \ref{fig:Layer_Ablation_Backdoor}. The backdoor attack affects each layer of the DNN when executed, and that effect is seen here with coverage being materially significant at each layer. Coverage is highest at residual block 1 and the final layer (76.9\% and 57.2\% respectively), for the outlier threshold of 0.01. 

\begin{figure}
    \centering
    \includegraphics[width=0.5\textwidth]{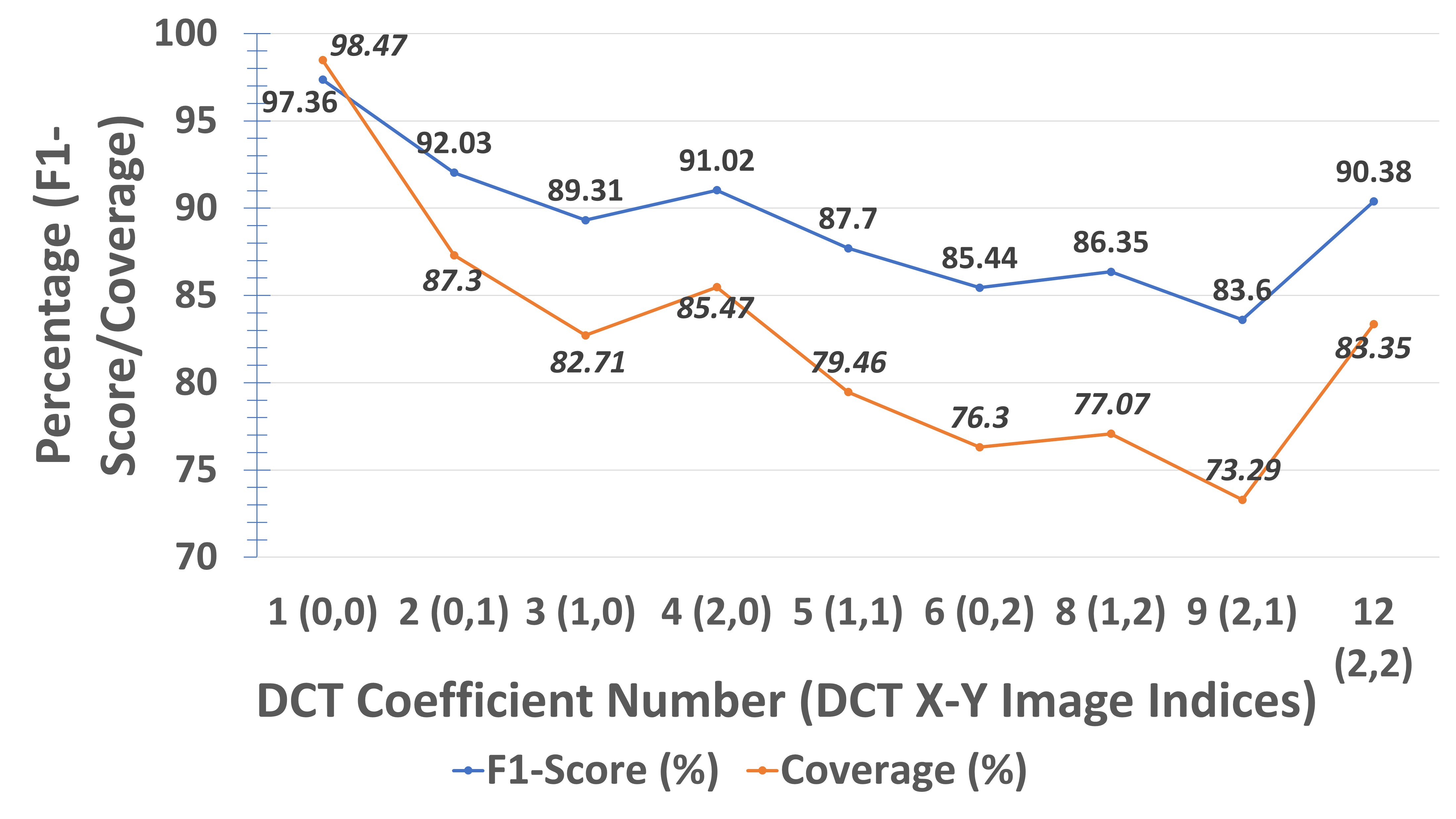}
    \caption{\textbf{\textit{Detection Against Input-Aware Backdoor on PreAct ResNet-18, Varying DCT Coefficient Chosen for PCA: }}The detector is connected concurrently on all four residual blocks and the final layer.}
    \label{fig:DCT_Ablation_Backdoor}
\end{figure}

(3) Detector performance for different DCT coefficients selected for the PCA is shown in Figure \ref{fig:DCT_Ablation_Backdoor}. Since $J=1$ here, changing the DCT coefficient affects performance. The italicized figures are the data point values for F1-score and the non-italicized ones are for the coverage. The DCT coefficient number is displayed along with its (X,Y) indices in the transformed image. The coverage drops sharply after the first DCT coefficient, with local peaks every fourth DCT coefficient (1, 4 and 8). The F1-score remains high due to the low false positive rate, but shows a similar trend.
\subsection{Adversarial Attack: Projected Gradient Descent}

The third attack tested is the Projected Gradient Descent (PGD) adversarial attack \cite{PGD}. The attacker is assumed to have information about the model gradients to compute the required input perturbations for the DNN to force a misclassification to any class. The PGD attack accomplishes this by attempting to find the input perturbations that maximize model loss while keeping perturbations smaller than an amount $\zeta$, measured as the $L_{\infty}$ norm of the difference between the clean input and perturbed input. This is done via a gradient descent taking steps in the direction of maximum loss, projecting the resultant perturbation back into the $L_{\infty}$ space around the clean input and iterating further. This is implemented for the clean CIFAR-10 ResNet-18 model used for the TBT attack via the Cleverhans v4.0.0 \cite{cleverhans} adversarial benchmarking library. The value of $\zeta$ of the PGD used here is 0.3. The attack forces misclassification with more than 99\% effectiveness for ResNet-18 on CIFAR-10.

The detector performance against the PGD attack is tested on ResNet-18 for cases 1-3 of Section \ref{networks}.

\begin{figure}
    \centering
    \includegraphics[width=0.5\textwidth]{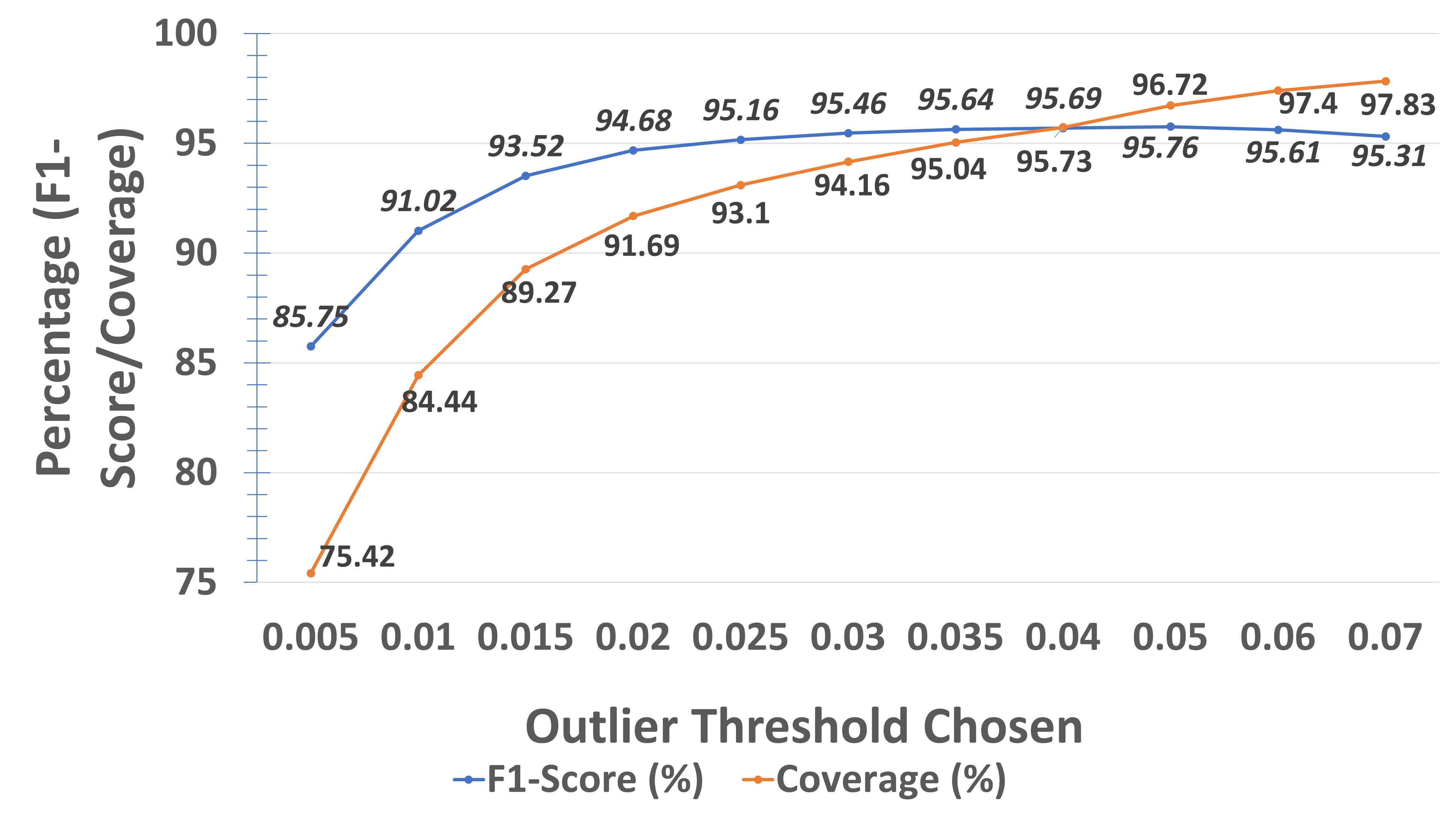}
    \caption{\textbf{\textit{Detection Performance Against PGD Attack on ResNet-18, Varying Thresholds: }}The detector is connected concurrently on all four residual blocks and the final linear layer.}
    \label{fig:Thresh_PGD}
    \vspace{-0.2cm}
\end{figure}

(1) Detector performance for different levels of outlier detection thresholds is shown in Figure \ref{fig:Thresh_PGD}. The F1-score peaks at $\varepsilon=0.04$. This gives a coverage of 95.73\% and a false positive rate of 4.35\%. $\varepsilon=0.03$ yields a coverage of 94.16\% and a false positive rate of 3.1\%. The italicized figures are the data point values for F1-score and the non-italicized ones are for the coverage.

\begin{figure}
    \centering
    \includegraphics[width=0.5\textwidth]{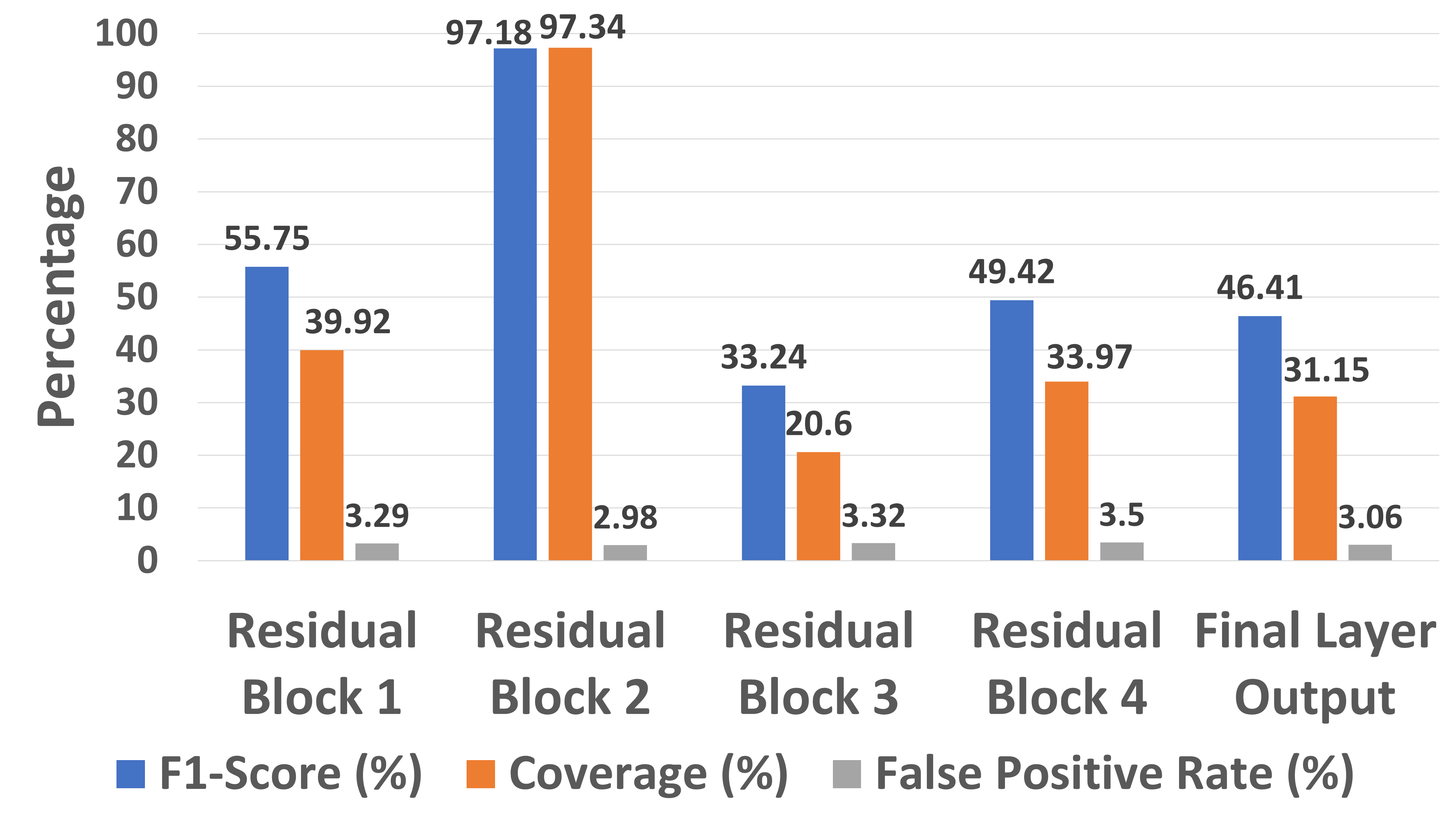}
    \caption{\textbf{\textit{Detection Performance By Layer Against PGD Attack: }}The detector is connected across each residual block and the final layer of ResNet-18 \textit{individually}}
    \label{fig:Layer_Ablation_PGD}
    \vspace{-0.2cm}
\end{figure}

(2) Detector performance for each residual block is shown in Figure \ref{fig:Layer_Ablation_Backdoor}. The backdoor attack is not targeted at any one layer, thus coverage is significant at each layer. Coverage is highest at residual block 2 (76.9\% and a 2.98\% false positive Rate), for the outlier threshold of 0.03. 

\begin{figure}
    \centering
    \includegraphics[width=0.5\textwidth]{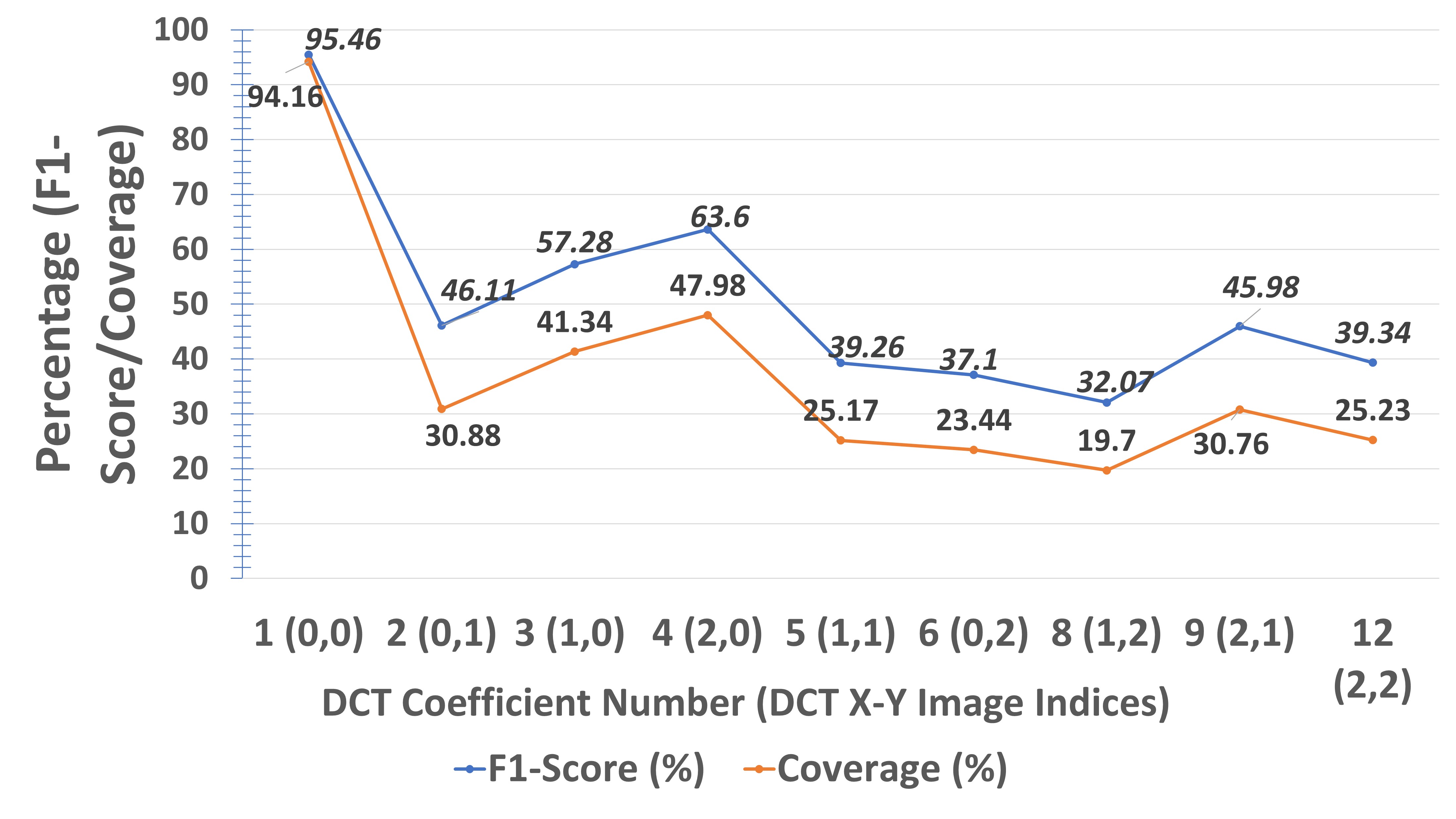}
    \caption{\textbf{\textit{Detection Performance Against PGD Attack, Varying DCT Coefficients Chosen for PCA: }}The detector is connected concurrently on all four residual blocks and the final linear layer.}
    \label{fig:DCT_Ablation_PGD}
\end{figure}

(3) Detector performance for different DCT coefficients selected for the PCA is shown in Figure \ref{fig:DCT_Ablation_PGD}. The italicized figures are the data point values for F1-score and the non-italicized ones are for the coverage. The DCT coefficient number is displayed along with its (X,Y) indices in the transformed image. The coverage and F1-score drop sharply after the first DCT coefficient.

\subsection{Algorithm Overhead}\label{HardwareOverhead}

\begin{table}[]
\centering
\caption{\textbf{Detector Performance Overhead} when connected concurrently across all four residual blocks and the final layer}
\vspace{0.1in}
\begin{tabular}{|l|l|l|}
\hline
\textbf{Metric}                                                                                               & \begin{tabular}[c]{@{}l@{}}\textbf{Overhead (\%)} \\ \textbf{(PreAct} \\ \textbf{ResNet-18)}\end{tabular} & \begin{tabular}[c]{@{}l@{}}\textbf{Overhead (\%)} \\ \textbf{(ResNet-18)}\end{tabular} \\ \hline
\begin{tabular}[c]{@{}l@{}}Floating Point \\ Operations (Single \& \\ Double Precision)\end{tabular} & 1.31                                                                           & 1.13                                                                 \\ \hline
\begin{tabular}[c]{@{}l@{}}Vector Operations \\ (Single \& Double \\ Precision)\end{tabular}         & 19.17                                                                          & 16.02                                                                \\ \hline
\begin{tabular}[c]{@{}l@{}}Cache Reads \\ (L2 and L3 Caches)\end{tabular}                            & 175.8                                                                         & 117.8                                                               \\ \hline
\begin{tabular}[c]{@{}l@{}}Cache Writes \\ (L2 and L3 Caches)\end{tabular}                           & 1.478                                                                          & 0.92                                                                \\ \hline
\end{tabular}
\label{overheadtable}
\end{table}

The performance overhead of TESDA relative to the DNN when connected across all residual blocks and the final linear layer is shown in Table \ref{overheadtable}. PreAct ResNet-18 uses the trained model provided by the authors of \cite{InputAwareBackdoor}. ResNet-18 uses the trained model provided by the authors of \cite{TBT}. Overhead was calculated using the Performance Application Programming Interface library \cite{PAPI}, which allows access to CPU counters. The average overhead was calculated across the test data of GTSRB for PreAct ResNet-18 and CIFAR-10 for ResNet-18. The DNNs were written in PyTorch \cite{pytorch}. The detector used Scikit-Learn \cite{sklearn}. This was run on an Intel Xeon W-2123 CPU. The counters were run for the DNN across DNN inference alone. The counters for the detector were run across conversion of intermediate layer data to suitable formats, feature extraction and outlier detection on the extracted $\theta$.

The overhead is ultra-low for floating point operations, at around 1\%. Vector operations are still very low at less than 20\% overhead. Since the detector does not write much to cache and reads a great deal from the cache (layer outputs for analysis), the cache write overhead is extremely low (0.8-1.2\%). The cache read overhead is high compared to the DNN, which does not read much data from the CPU cache. The detector has low overhead without requiring dedicated hardware, run using common libraries on a general purpose CPU. Overhead can be reduced by reducing the number of layers the detector takes inputs from concurrently.

\section{Ablation Studies and Discussion}\label{Discussion}
\subsection{Threshold Variation}

As expected, increasing the outlier threshold $\varepsilon$ (Figures \ref{fig:TBT_thresh}, \ref{fig:Thresh_Backdoor} and \ref{fig:Thresh_PGD}) increases detection and false positive rate. Assuming a general detection scenario with the detector built pre-deployment, this would give coverages of 91.69\% on PGD attacks and 99\% on TBT for a false positive rate of 1.99\% for ResNet-18 on CIFAR-10 with $\varepsilon$ at 0.02. The TBT is seen to achieve near-total detection coverage at very low thresholds, indicative of the drastic change in behavior in the final (targeted) layer caused by the attack. More subtle attacks like the input-aware backdoor and PGD are less extreme and therefore have lower coverage at low thresholds.

\subsection{Layer Ablations}

Each of the attacks tested shows different behavior with the detector restricted to one layer (Figures \ref{fig:TBT_Layer_Ablation}, \ref{fig:Layer_Ablation_Backdoor} and \ref{fig:Layer_Ablation_PGD}). The \textit{TBT} attacks the final layer weights and thus shows high coverage in that layer alone. The \textit{input-aware backdoor} attack shows high coverage across the output of Residual Block 1 and the final layer, likely due to the trigger patterns changing the behavior of Residual Block 1 more than succeeding units. The final layer behavior changes due to earlier layers being affected by the attack. The \textit{PGD} attack is most easily detected at the output of Residual Block 2 ($>90\%$ coverage). This residual block can be considered the diverging point \cite{ExplainableAdversarial} where the adversarial attack forces misclassification. From the experimental results, it can be seen that Residual Blocks 3 and 4 are of lower significance for all considered attacks. The lightest possible detector configuration for these attacks would thus be connected across Residual Blocks 1, 2 and the final layer.

\subsection{DCT Coefficient Variation}

Variation of DCT coefficients chosen from each convolutional layer's output to check effect on detection was done for the input-aware backdoor attack on PreAct ResNet-18 and the PGD attack on ResNet-18 (Figures \ref{fig:DCT_Ablation_Backdoor} and \ref{fig:DCT_Ablation_PGD}). For the input-aware backdoor attack, detection remains materially significant across a range of DCT coefficients' final $\alpha$ values. The backdoor's alteration of the features extracted by the DNN thus affects a range of frequencies. For the \textit{PGD} attack, the constraint to limit the deviation in the image based on the $L_{\infty}$ norm of the perturbation seems to limit the effect of the PGD attack on higher frequencies. It instead appears to add perturbations that shift the final PCA coefficient of the first (DC) DCT coefficient. The use of \textit{more than one} DCT coefficient has little impact on detection, marginally raising coverage for the input-aware backdoor. Further details can be found in Appendix \ref{appendix_dct_coeffs}.

\section{Conclusion}\label{Conclusion}

Our work presents and validates TESDA, a scheme for DNN security attack detection on two different types of state of the art attacks.
We find that it is extremely comparable in terms of performance to strong baselines from prior art, while operating at a much lower hardware overhead and working for a broader range of attacks.
The extension of TESDA to online \emph{defense} along with further study of its real, in-field performance is envisioned as future work.

\section*{Acknowledgements}
This research was supported by the Semiconductor Research Corporation under Auto Task 2892.001 and in part by the U.S. National Science Foundation under Grant S\&AS:1723997.

\bibliography{TESDA}
\bibliographystyle{mlsys2022}


\clearpage
\appendix
\section{Deriving a General Expression for $\Delta$ Dependent on $\varepsilon$}
\label{tuning_delta_details}

Given $\hat{\mu}, \hat{\Sigma}$ as our sample mean and covariance estimates, for a sample $\theta_i \in \mathbb{R}^k$ to be classified as an outlier we require that its squared Mahalanobis distance $d_i^2 \geq \Delta^2$.
Here, $d_i^2 = {(\theta_i-\hat{\mu})^T\hat{\Sigma}^{-1}(\theta_i-\hat{\mu})}$ and $\Delta$ is a specified threshold computed over the training set such that the number of outliers present in it are consistent with the provided estimate.
To that end, we know that no more than $m = \varepsilon n$ samples of our training set can have their $d_i^2 \geq \Delta^2$.
Assuming $d^2$ to be the random variable governing the squared Mahalanobis distances for over the training set, we have the inequality 
\begin{equation}
\label{eq1}
\mathbb{P}[d^2 \geq \Delta^2] \leq \varepsilon    
\end{equation}
Next, we note that the multivariate Chebyshev inequality in $k$ dimensions, with estimated mean and variance and over $h$ samples \cite{stellato2017multivariate} can be stated as
\begin{equation}
\label{eq2}
\mathbb{P}[d^2 \geq \delta^2] \leq \min\{1,\frac{k(h^2-1+h^2\delta^2)}{h^2 \delta^2}\}    
\end{equation}
We shall only consider the more interesting and non-trivial case where $\min\{1,\frac{k(h^2-1+h^2\delta^2)}{h^2 \delta^2}\} = \frac{k(h^2-1+h^2\delta^2)}{h^2 \delta^2}$ thus simplifying Eq. \ref{eq2} to
\begin{equation}
\mathbb{P}[d^2 \geq \delta^2] \leq \frac{k(h^2-1+h^2\delta^2)}{h^2 \delta^2} 
\end{equation}
Substituting $\delta^2 = \Delta^2$ and recalling from Section \ref{attack_detection} that when given $n$ as the total number of points in the training set, $h = \frac{n-k-1}{2} \approx \frac{n}{2}$ since $n \gg k$,  we finally get
\begin{equation}
\label{eq4}
\mathbb{P}[d^2 \geq \Delta^2] \leq \frac{k(n^2-4+2n\Delta^2)}{n^2 \Delta^2}   
\end{equation}
Equating Eq. \ref{eq1} and Eq. \ref{eq4}, we get
$$\varepsilon = \frac{k(n^2-4+2n\Delta^2)}{n^2\Delta^2} \Rightarrow \Delta = \sqrt{\frac{k(n^2-4)}{\varepsilon n^2 - 2nk}}$$

\section{Squared Mahalanobis Distances follow the $\chi^2_k$ Distribution }
\label{mahalanobis_chi_squared}

The squared Mahalanobis distance of a sample $\theta$ is
\begin{equation}\label{mahalanobis-squared-eq}
\begin{split}
d^2 = (\theta-\mu)^T\Sigma^{-1}(\theta-\mu)
\end{split}
\end{equation}
Since $\Sigma$ is symmetric positive definite, we rewrite Eq. \ref{mahalanobis-squared-eq} as
\begin{equation}\label{half_sigma_eq}
    d^2 = (\Sigma^{-\frac{1}{2}}(\theta-\mu))^T(\Sigma^{-\frac{1}{2}}(\theta-\mu))
\end{equation}
Setting $Y = \Sigma^{-\frac{1}{2}}(\theta-\mu)$ in Eq. \ref{half_sigma_eq}, $d^2 = Y^TY = ||Y||^2_2$
where $Y_i \sim N(0,1)\; \textrm{are i.i.d. } \forall i \in \{1,2,...,k\}$.

This makes the distribution for $d^2$ the same as that of the sum of squares of $k$ independent standard normal random variables, which by definition is the $\chi^2_k$ distribution.

\section{Sub-exponential Bound for $\Delta$}
\label{subexponential_delta_bounds}

The sub-exponential tail bound \cite{wainwright2019high} states that for a sub-exponential distribution $X$ with first moment $\mu$ and parameters $(\nu,b)$, 
\begin{equation}\label{subexp_bound_general}
\mathbb{P}[X \geq \mu + t] \leq \begin{cases}
\exp{(-\frac{t^2}{2\nu^2})} &0 \leq t \leq \frac{\nu^2}{b}\\
\exp{(-\frac{t}{2b})} &t > \frac{\nu^2}{b}
\end{cases}
\end{equation}

Using the facts that i) $d^2 \sim \chi^2_k$\; ii) $\chi^2_k$ is sub-exponential with parameters ($2k,4$) \cite{ghosh2021exponential} and iii) $\mu = k$ for $\chi_k^2$

We can write
\begin{equation}\label{subexp_bound_specific}
\mathbb{P}[{d}^2 \geq \Delta^{2}] \leq \begin{cases}
\exp{(-\frac{(\Delta^2-k)^2}{8k^2})} &\sqrt{k} \leq \Delta \leq \sqrt{k^2+k}\\
\exp{(-\frac{\Delta^2-k}{8})} &\Delta > \sqrt{k^2+k}
\end{cases}
\end{equation}
where $\Delta^2 = \mu + t \Rightarrow t = \Delta^2 - \mu$.

Comparing Eq. \ref{subexp_bound_specific} with $\mathbb{P}[d^2 \geq \Delta^2] \leq \varepsilon$ we have the following two cases:

\begin{enumerate}
    \item \textbf{$\sqrt{k} \leq \Delta \leq \sqrt{k^2+k}$}
    $$\varepsilon = \exp{(-\frac{(\Delta^2-k)^2}{8k^2})} \Rightarrow \Delta = \sqrt{4k\sqrt{\ln(\frac{1}{\varepsilon^2})}+k}$$
    \item \textbf{$\Delta > \sqrt{k^2+k}$}
    $$\varepsilon = \exp{(-\frac{\Delta^2-k}{8})} \Rightarrow \Delta = \sqrt{8\ln{(\frac{1}{\varepsilon})}+k}$$
\end{enumerate}

We note that even though these bounds do not cover the case where $0 \leq \Delta \leq \sqrt{k}$, for the purposes of outlier detection the value of $\Delta > \mu\;(= k)$ by definition.

\section{Chernoff Style Bound for $\Delta$}
\label{chernoff_delta_bounds}

The Chernoff bound \cite{wainwright2019high} is a tail bound given by the inequality
\begin{equation}\label{chernoff_bound_eq}
\mathbb{P}[d^2 \geq t] \leq \exp(-\psi^*_x(t))    
\end{equation}
where $\psi^*_x(t)$ is defined as $\sup_{\lambda \geq 0}[\lambda t - \psi(\lambda)],$ and $\psi(\lambda)$ is the logarithm of the moment generating function of $d^2$.

Comparing Eq. \ref{chernoff_bound_eq} with $\mathbb{P}[d^2 \geq \Delta^2] \leq \varepsilon$ gives us
\begin{equation}\label{delta_squared_t_eq}
   \Delta^2 = t 
\end{equation}
As well as
$$
    \varepsilon = \exp(-\psi^*_x(t)) \Rightarrow \ln(\frac{1}{\varepsilon}) = \psi^*_x(t)
$$
Since $d^2 \sim \chi^2_k$ we know
\begin{equation}\label{log_mgf_chi_squared}
 \psi(\lambda) = -\frac{k}{2}\ln(1-2\lambda) 
\end{equation}
Subsequently, differentiating $t + \frac{k}{2}\log(1-2\lambda)$ w.r.t. $\lambda$ and setting it to $0$ gives
$$
    \lambda^* = \frac{t-k}{2t}
$$
Which when substituted in the defitintion of $\psi_x^*(t)$ yields
\begin{equation}\label{psi_star_eq}
\psi^*_x(t) = \frac{1}{2}(t-k + k\ln(k/t)) = \ln(1/\varepsilon)    
\end{equation}
Substituting Eq. \ref{delta_squared_t_eq} in Eq. \ref{psi_star_eq} and solving for $\Delta^2$ we get
$$
    \Delta^2 = -kW(\frac{-\varepsilon^{2/k}}{e}) \Rightarrow \Delta = \sqrt{-kW(\frac{-\varepsilon^{2/k}}{e})}
$$
where $W$ is the Lambert \emph{W} function \cite{bronstein2008algebraic}.

\section{Use of Multiple DCT Coefficients for Detection}\label{appendix_dct_coeffs}
\begin{figure}[h]
    \centering
    \includegraphics[width=0.5\textwidth]{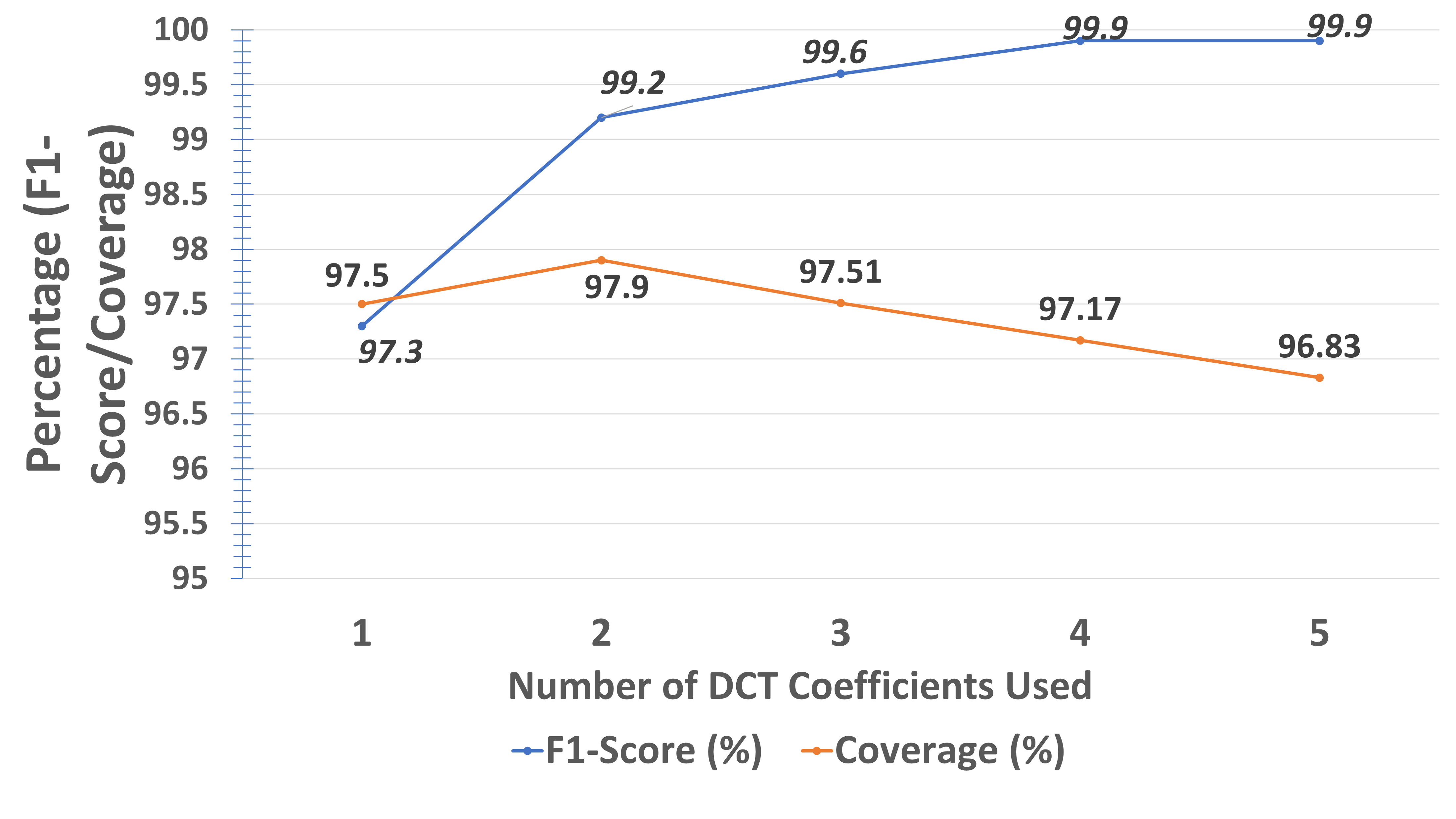}
    \caption{Detector performance for input-aware backdoor attacks on PreAct ResNet-18 using the first $k$ DCT coefficients, $1\leq k\leq 5$. The detector is connected across all residual blocks and the final layer. $\varepsilon=0.005$ here. PCA and outlier detection are done in parallel for each DCT coefficient and detection is raised if \textit{any one} of their $\theta$s is an outlier.}
    \label{fig:appendix_dct_plot}
\end{figure}

The use of multiple DCT coefficients for detection was examined for PreAct ResNet-18 on GTSRB for the input-aware backdoor attack. The detector was connected across the four residual blocks of the DNN and the final layer concurrently and the first $k$ DCT coefficients were taken for detection, $1\leq k \leq 5$. The dimensionality reduction and feature extraction via PCA were done in parallel on each row of the DCT matrix $D_i$ of each layer $i$ as discussed in Section \ref{Details}. Outlier detection was done in parallel with an attack declared if \textit{any} of the $\theta$s from the DCT coefficients was an outlier with respect to its distribution using its own outlier detector. The results are shown in Figure \ref{fig:appendix_dct_plot}. The use of more DCT coefficients marginally raises detection coverage and lowers false positive rate at the cost of overhead increases from additional computation.

\end{document}